\documentclass[12pt]{article}

\usepackage[utf8]{inputenc}
\usepackage[T1]{fontenc}
\usepackage{authblk} 
\usepackage{amssymb}
\usepackage[a4paper, margin=2.5cm]{geometry}
\usepackage{fancyhdr}

\title{\textbf{Generation of metric waves from cause-effect processes}}

\author[1]{B. Paroli\thanks{\texttt{bruno.paroli@unimi.it}}}

\affil[1]{Dipartimento di Fisica, Universit$\acute{a}$ degli Studi di Milano - via G. Celoria, 16, 20133 Milano, Italy}

\date{\today} 

\begin{document}

\maketitle

\begin{abstract}

This work investigates the origin of field waves, explicitly dependent on distance, considering fundamental principles essentially based on cause-effect processes. The theoretical framework is applied to electromagnetic (EM) and gravitational (G) interactions, showing that the waves of the EM and G fields can be interpreted as being supported by waves of metric nature susceptible to perturbation. 
\end{abstract}

\section{Introduction}
Waves are fascinating phenomena in nature, their description involves two fundamental aspects: the perturbation of a physical quantity and its propagation. The methods that have contributed to the definition of the wave concept have an incredibly rich and extensive history, often representing complementary approaches to the description of wave phenomena.

Early observations concerning the propagation of acoustic waves can already be traced back to Aristotle (c. 350 BCE), who recognized the essential role of the medium in the transmission of sound \cite{Aristotele}. In Kepler studies, and particularly in the Harmonices Mundi (1619) \cite{Keplero}, the relationship between geometrical constructions, mathematical proportions, and harmonic relationships became a central element in the description of natural phenomena. Kepler work thus represents an important step towards the mathematical interpretation of harmonic phenomena, establishing a conceptual connection between geometry, mathematical ratios, and periodic celestial motions. This use of geometrical and mathematical relations in the description of natural phenomena has important antecedents in the works of Euclid \cite{Euclid} and Ibn al-Haytham \cite{Haytham1,Haytham}, whose geometrical treatments of optical phenomena provided an earlier mathematical framework for the study of light. 

The earliest comprehensive wave theories of light, addressing reflection, refraction, and related optical phenomena with remarkable mathematical and physical insight, are associated with the works of Huygens \cite{Huygens}, Young \cite{Young} and Fresnel \cite{Fresnel}. Alongside these developments in the study of light, the mathematical description of waves in differential form was developed by D'Alembert , who in his famous wave equation  \cite{Dalembert}, described waves through the local differential properties of a perturbation in space and time, providing a mathematical description of wave evolution and propagation. The mathematical structure of  D'Alembert equation subsequently became fundamental to the mathematical formulation of wave phenomena, including electromagnetic waves \cite{Maxwell}, and was later further explored in the context of general relativity \cite{Einstein} and quantum theories, notably in the Klein-Gordon equation \cite{Klein,Gordon}.

In this work, how the origin of waves is described starting from fundamental principles, regardless of the physical mechanisms and dynamics of the particular system. The proposed formulation is based on two simple assumptions involving cause-effect processes: i) The effect manifests itself at finite distance from the causal event. ii) The effect cannot manifest instantaneously at that distance. Both points i), ii) appear plausible in many real contexts, especially in the field of electric, magnetic and gravitational interactions whose fields propagates at limited speed. 

By denying instantaneity, the effect of a causal event cannot realize instantaneously at a finite distance from the event position, but it should necessarily manifest at a future time.
There should be a natural condition that allows the effect to manifest itself even when the causal event vanishes.

Therefore, we should expect in nature a general condition that allows us to distinguish the 'becoming' of the effect (also when the causal event vanishes) from its 'non-becoming' (for example due to the absence of the causal event at any time). Furthermore, such a condition should be independent of the physical nature that links the action of creating to the action of becoming and therefore exist for a wide class of phenomena.

In this work, a mathematical formulation is given in the form of theorem, proving that a generic wave of field can be seen supported by a wave of metric nature which involve only space and time. The source of such metric perturbation,  coincides with the source of the field suggesting that waves, regardless of the nature of the field, can be interpreted as the manifestation of the cause-effect process previously discussed. The proposed theorem, in its scalar formulation, shows the fundamental role of the distance: the wave propagation can be described by a wave of metric nature that, in the most relevant cases, is everywhere homogeneous except for the position of the source of the event.

\section{Theorem}

Let $F(D):\Omega'\subseteq \Re$ be a scalar field, of class $C^2$ in $\Omega'$ , generated by a source with retarded time position $\vec{s}(t')$, where $D(\vec{x},\vec{s})$ is the distance between $\vec{x}$ and $\vec{s}$, $\vec{x}$ is the space, $t$ and $t'$ are the time and the retarded time, respectively and let $f:\Omega\subseteq \Re^{n+2}\rightarrow \Re$ be of class $C^1$ in $\Omega$ such that, for $(\vec{x},t,t')\in \Omega$, $f(\vec{x},t,t')=vt'-vt+D(\vec{x},\vec{s}(t'))=0$ is an implicit form of the function $t'(\vec{x},t)$, where $v\neq0$ is the speed of the field propagation; then a wave of the field $F(D)$ is the composition of a metric wave $\left[\frac{\partial^2 D}{\partial \vec{x}^2}
-\frac{1}{v^2}\frac{\partial^2 D}{\partial t^2}\right]$ and a metric equation $\left[\frac{\partial D}{\partial \vec{x}}^2
-\frac{1}{v^2}\frac{\partial D}{\partial t}^2\right]$: 

\begin{equation}\label{result} 
\frac{\partial^2 F}{\partial \vec{x}^2}-
\frac{1}{v^2}\frac{\partial^2 F}{\partial t^2}
=\frac{\partial F}{\partial D}
\left[\frac{\partial^2 D}{\partial \vec{x}^2}
-\frac{1}{v^2}\frac{\partial^2 D}{\partial t^2}\right]
\end{equation}
$$
+\frac{\partial^2 F} {\partial D^2}
\left[\frac{\partial D}{\partial \vec{x}}^2
-\frac{1}{v^2}\frac{\partial D}{\partial t}^2\right],
$$

\noindent such that if\\

i) $D=||\vec{x}-\vec{s}||$ is the Euclidian unperturbed distance,

ii) $\gamma'$ is a geodesic of $D$ with $\vec{x} \in \gamma'$, $\vec{s} \in \gamma'$,

iii) the directional derivatives of the distance are evaluated along the geodesic direction,\\

\noindent then the metric wave is everywere homogeneous expect for $\vec{x}=\vec{s}$ and the source of the field is the source of the metric wave and takes the form,

\begin{equation}\label{result_2}
\frac{\partial^2 D}{\partial \vec{x}^2}
-\frac{1}{v^2}\frac{\partial^2 D}{\partial t^2}=2\delta(||\vec{x}-\vec{s}||)[K^2+3 K \partial_x \tilde{D}+1]\:,
\end{equation}

where $K=\frac{\partial t'}{\partial t} \frac{ds}{dt'} \frac{1}{v}$.\\

\noindent \textbf{Proof:} I start the proof estimating the ratios of retarded time elementary interval $\partial t'$  with respect to the space $\partial \vec{x}$ and the time $\partial t$ elementary intervals. Since $f$ is the implicit form of the differentiable function $t'(\vec{x},t)$, by using the Dini theorem \cite{Dini} we write

\begin{equation}\label{ratiox}
\frac{\partial t'}{\partial \vec{x}}=-\left[v+\frac{\partial D}{\partial t'}\right]^{-1} \frac{\partial \tilde{D}}{\partial \vec{x}},
\end{equation}

and

\begin{equation}\label{ratiot}
\frac{\partial t'}{\partial t}=\left[1+\frac{\partial D}{\partial t'}\gamma\right]^{-1}\:,
\end{equation}

where $\gamma=1/v$ and the tilded derivative $\frac{\partial \tilde{D}}{\partial \vec{x}}$ is intended for $\vec{s}$ independent of $\vec{x}$. Equations \ref{ratiox} and \ref{ratiot} show that 

\begin{equation}\label{property}
\frac{\partial t'}{\partial t}=
-v\frac{\partial t'}{\partial \vec{x}}
\left[\frac{\partial \tilde{D}}{\partial \vec{x}}\right]^{-1} \:.
\end{equation}

Since $F(D(\vec{x},t))$ is of class $C^2$ in $\Omega'$ we estimate the second derivatives in space and time. 
We write the partial derivatives of $F(D(\vec{x},t))$ as

\begin{equation}\label{FDt}
\frac{\partial}{\partial t}\frac{\partial F}{\partial D}=\frac{\partial^2 F}{\partial D^2} \frac{\partial D}{\partial t}=\frac{\partial^2 F}{\partial D^2}\frac{\partial D}{\partial t'}\frac{\partial t'}{\partial t}
\end{equation}

and

\begin{equation}\label{FDx}
\frac{\partial}{\partial \vec{x}}\frac{\partial F}{\partial D}=\frac{\partial^2 F}{\partial D^2} \frac{\partial D}{\partial \vec{x}}=\frac{\partial^2 F}{\partial D^2}\frac{\partial D}{\partial t'}\frac{\partial t'}{\partial \vec{x}} \:.
\end{equation}

We also write the spatial and temporal derivatives of  $F(D(\vec{x},t))$ as

\begin{equation}\label{firstx}
\frac{\partial F}{\partial \vec{x}}=\frac{\partial F}{\partial D}\frac{\partial D}{\partial \vec{x}}, 
\end{equation}

and

\begin{equation}\label{firstt}
\frac{\partial F}{\partial t}=\frac{\partial F}{\partial D}\frac{\partial D}{\partial t}\:. 
\end{equation}

The second spatial and temporal derivatives are then

\begin{equation}\label{secondx}
\frac{\partial^2 F}{\partial \vec{x}^2}=
\frac{\partial}{\partial \vec{x}}
\frac{\partial F}{\partial D}
\frac{\partial D}{\partial \vec{x}}+
\frac{\partial^2 D}{\partial \vec{x}^2}
\frac{\partial F}{\partial D}
\end{equation}

and

\begin{equation}\label{secondt}
\frac{\partial^2 F}{\partial t^2}=
\frac{\partial}{\partial t}
\frac{\partial F}{\partial D}
\frac{\partial D}{\partial t}+
\frac{\partial^2 D}{\partial t^2}
\frac{\partial F}{\partial D},
\end{equation}

by using equations \ref{secondx} and \ref{secondt} we write

\begin{equation}\label{onde_1}
\frac{\partial^2 F}{\partial \vec{x}^2}-
\frac{1}{v^2}\frac{\partial^2 F}{\partial t^2}=
\frac{\partial}{\partial \vec{x}}
\frac{\partial F}{\partial D}
\frac{\partial D}{\partial \vec{x}}+
\frac{\partial^2 D}{\partial \vec{x}^2}
\frac{\partial F}{\partial D}
\end{equation}
$$
-\frac{1}{v^2} \frac{\partial}{\partial t}
\frac{\partial F}{\partial D}
\frac{\partial D}{\partial t}-
\frac{1}{v^2}\frac{\partial^2 D}{\partial t^2}
\frac{\partial F}{\partial D}\:,
$$

by using Eq.s \ref{FDt}, \ref{FDx} we write

\begin{equation}\label{onde}
\frac{\partial^2 F}{\partial \vec{x}^2}-
\frac{1}{v^2}\frac{\partial^2 F}{\partial t^2}
=\frac{\partial F}{\partial D}
\left[\frac{\partial^2 D}{\partial \vec{x}^2}
-\frac{1}{v^2}\frac{\partial^2 D}{\partial t^2}\right]
\end{equation}
$$
+\frac{\partial^2 F} {\partial D^2}
\left[\frac{\partial D}{\partial \vec{x}}^2
-\frac{1}{v^2}\frac{\partial D}{\partial t}^2\right],
$$

and in compact notation as

\begin{equation}\label{onde_c}
\partial_{xx}F-\frac{1}{v^2}\partial_{tt}F=
\partial_{D}F \left[ \partial_{xx} D - \frac{1}{v^2} \partial_{tt}D\right]
\end{equation}
$$
+\partial_{DD}F \left[\partial_x D^2-\frac{1}{v^2}\partial_t D^2 \right] \:.
$$


Since $D=||\vec{x}-\vec{s}||$ is the euclidian distance between the source position $\vec{s}(t'(\vec{x},t))$ and the space position $\vec{x}$, the first derivatives of $D$ are written as

\begin{equation}\label{space_D}
\frac{\partial D}{\partial \vec{x}}= \hat{r} \cdot \hat{n}-\hat{r} \cdot \hat{m} \frac{ds}{dt'}\frac{\partial t'}{\partial \vec{x}}
\end{equation} 

and

\begin{equation}\label{time_D}
\frac{\partial D}{\partial t}= -\hat{r} \cdot \hat{m} \frac{ds}{dt'}\frac{\partial t'}{\partial t} ,
\end{equation} 

where $\hat{r}=\frac{\vec{x}-\vec{s}}{D}$, $\hat{n}$, $\hat{m}(\hat{n})$ are unit vectors and $\frac{ds}{dt'}=\frac{d\vec{s}}{dt'}\cdot \hat{m}$. The unit vector $\hat{m}(\hat{n})$ identifies the direction of the displacement $d\vec{s}$ along $\hat{m}$ due to the displacement $\partial\vec{x}$ along $\hat{n}$.

Using Eq. \ref{space_D} and \ref{time_D}, the metric wave $\partial_{xx} D - \frac{1}{v^2} \partial_{tt}D$ is written in compact form as

\begin{equation}\label{metric}
\partial_x (\hat{r} \cdot \hat{n})-\partial_x (\hat{r} \cdot \hat{m}) d_{t'}s\:\partial_x t'+\frac{1}{v^2} \partial_t (\hat{r} \cdot \hat{m})d_{t'}s\:\partial_t t'
\end{equation} 
$$
-\partial_x (d_{t'}s\:\partial_x t')(\hat{r} \cdot \hat{m})+\frac{1}{v^2}\partial_t (d_{t'}s\:\partial_t t')(\hat{r} \cdot \hat{m}),
$$

while, by substituting Eq. \ref{ratiox} in Eq. \ref{space_D}, the metric equation $\partial_x D^2-\frac{1}{v^2}\partial_t D^2$ is written as 

\begin{equation}\label{non-linear}
K^2 (\hat{r} \cdot \hat{m})^2 (\partial_x \tilde{D}^2-1)+2K (\hat{r} \cdot \hat{n})(\hat{r} \cdot \hat{m}) \partial_x \tilde{D}+ (\hat{r} \cdot \hat{n})^2
\end{equation} 

where $K=\frac{d_{t'}s\:\partial_t t'}{v}$.


Since $\gamma'$ is the geodesic of the distance $D(\vec{x},\vec{s})$ such that $\vec{x} \in \gamma'$, $\vec{s} \in \gamma'$ and the directional derivatives of the distance are evaluated along the geodesic direction: $\hat{m}=\hat{n}$, $|\hat{r} \cdot \hat{n}|=1$, therefore we write

\begin{equation}\label{cost_x_n}
\partial_x (\hat{r} \cdot \hat{n})=2 \delta(||\vec{x}-\vec{s}||) (1-d_{t'}s\:\partial_x t')
\end{equation} 

\begin{equation}\label{cost_x_m}
\partial_x (\hat{r} \cdot \hat{m})=2 \delta(||\vec{x}-\vec{s}||) (1-d_{t'}s\:\partial_x t') 
\end{equation}

and

\begin{equation}\label{cost_s}
\partial_s (\hat{r} \cdot \hat{m})=-2 \delta(||\vec{x}-\vec{s}||)\:.
\end{equation} 

By substituting Eq.s \ref{cost_x_n}, \ref{cost_x_m} and \ref{cost_s} into Eq. \ref{metric} and since

\begin{equation}\label{prop}
\partial_t (\hat{r} \cdot \hat{m})=\partial_s (\hat{r} \cdot \hat{m}) d_{t'}s\: \partial_t t'
\end{equation} 

we write 

\begin{equation}\label{geodesic}
\partial_{xx}D-\frac{1}{v^2}\partial_{tt}D=2 \delta(||\vec{x}-\vec{s}||) (1-d_{t'}s\:\partial_x t')^2
\end{equation}
$$
-\frac{1}{v^2}2 \delta(||\vec{x}-\vec{s}||) (d_{t'}s\:\partial_t t')^2
$$
$$
-\partial_x (d_{t'}s\:\partial_x t')(\hat{r} \cdot \hat{m})+\frac{1}{v^2}\partial_t (d_{t'}s\:\partial_t t')(\hat{r} \cdot \hat{m}),
$$

by using Eq. \ref{property} we write

\begin{equation}\label{1-D_last}
\partial_{xx}D-\frac{1}{v^2}\partial_{tt}D=
\end{equation}
$$
+2 \delta(||\vec{x}-\vec{s}||)[1+2\partial_{t'} s\gamma \partial_t {t'} \partial_x \tilde{D} +\gamma^2 \partial_x \tilde{D}^2(\partial_{t'}s \partial_t {t'})^2]-\frac{1}{v^2} 2\delta(||\vec{x}-\vec{s}||)(\partial_{t'}s \partial_t {t'})^2
$$
$$
-\partial_x (d_{t'}s\:\partial_x t')(\hat{r} \cdot \hat{m})+\frac{1}{v^2}\partial_t (d_{t'}s\:\partial_t t')(\hat{r} \cdot \hat{m})
$$

or

\begin{equation}\label{1-D_last_K}
\partial_{xx}D-\frac{1}{v^2}\partial_{tt}D=2 \delta(||\vec{x}-\vec{s}||)+4 \delta(||\vec{x}-\vec{s}||) K \partial_x \tilde{D}
\end{equation}
$$
-(\hat{r} \cdot \hat{m})[\partial_x (d_{t'}s\:\partial_x t')-\frac{1}{v^2}\partial_t (d_{t'}s\:\partial_t t')]\:.
$$

The second row of Eq. \ref{1-D_last_K} can be written as (see appendix for derivation)

\begin{equation}\label{1-D_last_K_new}
-(\hat{r} \cdot \hat{m})[\partial_x (d_{t'}s\:\partial_x t')-\frac{1}{v^2}\partial_t (d_{t'}s\:\partial_t t')]=
\end{equation}
$$
-\partial_x \tilde{D}[-2 \delta(||\vec{x}-\vec{s}||))K^2 \partial_x \tilde{D}-2 \delta(||\vec{x}-\vec{s}||)K]
$$

and hence

\begin{equation}\label{1-D_last_K_new_new}
\partial_{xx}D-\frac{1}{v^2}\partial_{tt}D=2 \delta(||\vec{x}-\vec{s}||)+6 \delta(||\vec{x}-\vec{s}||) K \partial_x \tilde{D}+2\delta(||\vec{x}-\vec{s}||)K^2
\end{equation}
$$
=2\delta(||\vec{x}-\vec{s}||)[K^2+3 K \partial_x \tilde{D}+1]\:.
$$

The metric wave is everywere homogeneous expect for $\vec{x}=\vec{s}$ and the source of the field is the source of the metric wave. \hfill $\blacksquare$

\section{Discussion}

The first result shown in Eq. 1 highlight, under the assumption of the initial hypothesis, as a field wave explicitly dependend on distance, can be represented as a superposition of a purely metric wave and a metric equation proportional to the curvature of the field in the space. The metric wave is susceptible to perturbation, and its disturbance radiates and propagates as a wave.

As shown in Eq. 2, the metric wave is homogeneous for $\vec{x} \neq \vec{s}$, and is thus formally capable of self-sustaining and propagating even in the absence of the field source (for example due to the source annihilation). Moreover, the metric wave involves only spatial and temporal quantities $(\vec{s}, t')$ and $(\vec{x}, t)$ i.e. two well-defined events namely, the cause and the effect. Equation 2 shows that the source of the field coincides with the source of the metric wave, highlighting how the metric itself (and not just the field) becomes central in the manifestation of the wave. The fundamental origin of the metric waves described here lies in the inability of the causal event to become an immediate effect, thereby involving and perturbing the metric that connects the cause to the effect (from the event $(\vec{s}, t')$ to event $(\vec{x}, t)$). 

\subsection{Cause-effect waves from symmetric pulsations}

On the basis of the previous results, it becomes possible to predict or reinterpret phenomena by relying on very fundamental concepts: specifically, the manifestation of an effect generated by its corresponding cause as the origin of a wave. For instance, metric waves can also be generated by symmetric pulsations (i.e. oscillating monopoles), in apparent contradiction with the Birkhoff theorem \cite{Birkhoff,Pappas}. A careful analysis shows that the contradiction is only apparent. In fact, Birkhoff theorem states that symmetric pulsations give rise to static solutions; therefore, the cause does not produce any effect, in agreement with the proposed theorem. There is, however, an exception predicted by the proposed theory, which allows for the generation of monopole waves without violating the Birkhoff theorem. If the pulsation involves the creation and annihilation of mass, while still maintaining purely spherical symmetry (at least during the existence of the particle), then the cause (i.e. the variation in mass) would indeed produce an effect (i.e. a variation in the gravitational field). Unfortunately, waves of pure gravitational nature arising from creation/annihilation processes would be extremely difficult to observe (since creating and destroying masses capable of producing experimentally detectable waves is a major challenge). An exceptional consequence of the creation/annihilation waves is that particle acceleration is not necessarily required, at least in principle, to generate cause-effect waves. Another interesting example is the  oscillating electric dipole. The motion of two opposite charges also implies motion of their respective masses. According to the proposed theory, this constitutes a cause that gives rise to a corresponding delayed effect, which would necessarily lead to the generation and propagation of a metric wave. 

\section{Conclusions}

In this work, a class of waves generated by the delayed manifestation of an effect is presented in the form of a theorem.
Specifically, the theorem is based on two main results. The first result (Eq. 1) shows how field waves explicitly dependent on distance can be described by means of metric waves that support the propagation of the field and a metric equation that describes how the field evolves as it propagates through the space.

The second result (Eq. 2) demonstrates that metric waves are, in all respects, propagating waves that are homogeneous everywhere except at the field source position. These results can be applied, for example, to electric, magnetic, and gravitational fields. 

In the case of electric and magnetic fields, although metric waves are compatible with independent propagation of the electric and magnetic components separately, any time variation in one component necessarily leads to the generation of the other component generating electromagnetic radiation. 
 
In the case of gravitational waves, it is natural to ask whether metric waves correspond to gravitational waves, since both involve perturbations of space. This issue is addressed by the second result of the theorem (see Eq. 2), that generally shows how metric waves are capable of propagating and transferring energy over long distances, even if not necessarily described in the framework of general relativity. Euclidean metric is considered in this work which is valid for small perturbations of the spacetime curvature, were space and time are treated here separately. 

\section{Appendix}

Let $\vec{s}(t')$ be the source position at the retarded time $t'$ then

\begin{equation}\label{appendix_2}
\partial_x (d_{t'}s\:\partial_x t')+\frac{1}{v^2}\partial_t (d_{t'}s\:\partial_t t')=-2\delta(||\vec{x}-\vec{s}||) K^2 \partial_x\tilde{D}-2\delta(||\vec{x}-\vec{s}||)K\:.
\end{equation}

\noindent \textbf{Proof:}
\begin{equation}\label{appendix_3}
\partial_x (d_{t'}s\:\partial_x t')+\frac{1}{v^2}\partial_t (d_{t'}s\:\partial_t t')=d_{t'}s\:[\partial_x(\partial_x t')-\frac{1}{v^2}\partial_t(\partial_t t')]
\end{equation}

the quantity between the square brackets can be written as 

\begin{equation}\label{appendix_4}
\partial_x(\partial_x t')-\frac{1}{v^2}\partial_t(\partial_t t')=
-\partial_x(\partial_t t')\gamma \partial_x \tilde{D}-\gamma \partial_t t' \partial_x (\partial_x \tilde{D})-\frac{1}{v^2}\partial_t (\partial_t t'),
\end{equation}

while the quantities $\partial_x(\partial_t t')$ and $\partial_t(\partial_t t')$ can be explicitly written as

\begin{equation}\label{appendix_5}
\partial_x(\partial_t t')=2\delta(||\vec{x}-\vec{s}||)\gamma \partial_t t'^2 d_{t'}s\:+2\delta(||\vec{x}-\vec{s}||)d_{t'}s\:^2 \gamma^2 \partial_t t'^3 \partial_x \tilde{D}-|d_{t't'}\vec{s}| \partial_t t'^3 \gamma^2
\end{equation}

and

\begin{equation}\label{appendix_6}
\partial_t(\partial_t t')=-2\delta(||\vec{x}-\vec{s}||)\gamma d_{t'}s\:^2 \partial_t t'^3+\gamma |d_{t't'}\vec{s}|\partial_t t'^3 \partial_x \tilde{D}\:. 
\end{equation}

By substituting Eq. \ref{appendix_5} and Eq. \ref{appendix_6} in Eq. \ref{appendix_4} and using Eq. \ref{appendix_3} we obtain

\begin{equation}\label{appendix_7}
\partial_x (d_{t'}s\:\partial_x t')+\frac{1}{v^2}\partial_t (d_{t'}s\:\partial_t t')=-\gamma^2 2\delta(||\vec{x}-\vec{s}||) \partial_t t'^2 d_{t'}s\:^2 \partial_x \tilde{D}-\gamma \partial_t t' d_{t'}s\: 2 \delta(||\vec{x}-\vec{s}||)
\end{equation}

being 

\begin{equation}\label{appendix_8}
K=\gamma \partial_t t' d_{t'}s
\end{equation}

we write,

\begin{equation}\label{appendix_9}
\partial_x (d_{t'}s\:\partial_x t')+\frac{1}{v^2}\partial_t (d_{t'}s\:\partial_t t')=-2\delta(||\vec{x}-\vec{s}||) K^2 \partial_x\tilde{D}-2\delta(||\vec{x}-\vec{s}||)K\:.
\end{equation}

\hfill $\blacksquare$

\section*{Acknowledgements}
This work is dedicated to Gabriel.


\begin{thebibliography}{9}

\bibitem{Aristotele}
Aristotele,
\textit{De Anima},
II, 8, 419b--420a,
c. 350 BCE.

\bibitem{Keplero}
J. Kepler,
\textit{Harmonices Mundi},
Linz, 1619.

\bibitem{Euclid}
Euclid,
\textit{Optica},
c. 300 BCE.

\bibitem{Haytham1}
A. Ibn al-Haytham,
\textit{Kit\=ab al-Man\=a\d{z}ir},
c.~1011--1021.

\bibitem{Haytham}
A. Ibn al-Haytham,
\textit{The Optics of Ibn al-Haytham. Books I--III: On Direct Vision},
trad. A. I. Sabra,
London, Warburg Institute, University of London, 1989.

\bibitem{Huygens}
C. Huygens,
\textit{Trait\'e de la lumi\'ere},
Leiden, 1690.

\bibitem{Young}
T. Young,
\textit{The Bakerian Lecture: On the Theory of Light and Colours},
Philosophical Transactions of the Royal Society of London,
vol. 92, 12--48, 1802.

\bibitem{Fresnel}
A. Fresnel,
\textit{M\'emoire sur la diffraction de la lumi\'ere},
M\'emoires de l'Acad\'emie royale des sciences de l'Institut de France,
vol. V, 339--475, 1826.

\bibitem{Dalembert}
J. d'Alembert,
\textit{Recherches sur la courbe que forme une corde tendue mise en vibration},
Histoire de l'Acad\'emie royale des sciences et des belles-lettres de Berlin,
vol. III, 214--219, 1747.

\bibitem{Maxwell}
J. C. Maxwell,
\textit{A Dynamical Theory of the Electromagnetic Field},
Philosophical Transactions of the Royal Society of London,
vol. 155, 459--512, 1865.

\bibitem{Einstein}
A. Einstein,
\textit{N\"aherungsweise Integration der Feldgleichungen der Gravitation},
Sitzungsberichte der K\"oniglich Preu$\beta$ischen Akademie der Wissenschaften, 688--696, 1916.

\bibitem{Klein}
O. Klein,
\textit{Quantentheorie und f\"unfdimensionale Relativit\"atstheorie},
Zeitschrift f\"ur Physik,
vol. 37, 895--906, 1926.

\bibitem{Gordon}
W. Gordon,
\textit{Der Comptoneffekt nach der Schr\"odingerschen Theorie},
Zeitschrift f\"ur Physik,
vol. 40, 117, 1926.

\bibitem{Dini}
U. Dini,
\textit{Fondamenti per la teorica delle funzioni di variabili reali},
Pisa, Tipografia T. Nistri, 1878.

\bibitem{Birkhoff}
G. D. Birkhoff,
\textit{Relativity and Gravitation},
Proceedings of the National Academy of Sciences,
vol. 9, 233--241, 1923.

\bibitem{Pappas}
R. C. Pappas,
\textit{Proof of ``Birkhoff's theorem'' in electrodynamics},
American Journal of Physics,
vol. 52, 3, 255--256, 1984.







\end{thebibliography}
\end{document}